# A low-cost mirror mount control system for optics setups

Maithreyi Gopalakrishnan[1,2] and Markus Gühr[1*]

[1]PULSE Institute, SLAC National Accelerator Laboratory, 2575 Sand Hill Road, Menlo Park, CA 94025
[2]University of Colorado, Boulder, 552 UCB, Boulder, CO 80309-0552

*mguehr@stanford.edu

We describe a flexible, simple to build, low-cost, and computer-controlled optical mirror actuator system we developed for undergraduate research laboratories. Geared motors for hobby robotics are controlled by an Arduino microcontroller board in combination with an H bridge to finely position the mirror mount actuators. We present a graphical user interface based on the free Python script language. The price of the fully controlled actuator system is thus only a small fraction of the price of any commercial system; however, it is quickly implementable due to the use of open hardware electronics. We show the performance of the system and give an outlook for future expansions and use in advanced optical setups.

## I. Introduction

Manipulating beams of light is one of the fundamental technical tasks in optical setups. The simplest and most used means to accomplish this uses mirrors, which are mostly aligned by hand. Commercial solutions for remote/computer control of mirror mounts are available and utilized for special purposes. In environments with restricted access, for example in combination with x-rays at synchrotrons [1] or free electron lasers [2-4], optical beams are steered from outside the restricted area. In addition, beam pointing stabilization can be accomplished by combining actuated and computer controlled mirror mounts with position sensitive devices [5]. The objective of this research is to develop a robust, easily available, and inexpensive alternative to highly developed commercial solutions in order to find wide diffusion in undergraduate research laboratories. This general approach has recently been facilitated due to novel open hardware solutions [6]. These include open microcontroller boards such as the Arduino, which has recently found applications in many scientific disciplines, from material science [7] and optics, [8] to neuroscience [9]. Due to the straightforward implementation and low cost, our motorized and computer controlled beam steering can be used in setups where their implementation has been too costly and/or laborious in the past. A large variety research from undergraduate projects to work at major research facilities can benefit from our effort.

Our setup is divided into three main parts: software, electronics, and mechanics. The user interface resides on the computer and is implemented in the Python script language. This interface sends data to an Arduino microcontroller board via a USB interface. The electronics, consisting of the Arduino and an H bridge, create electrical pulses which drive the motors on the mirror mount. The mechanics consists of motors, which clamp on a commercial mirror mount in a flexible way.

In the following we will show the mechanical, electronic and software basics of the system. We characterize the backlash of the system and show its angular stability and sensitivity. The GUI and Arduino program code are provided in the appendix.

## II. Setup

Below we describe the setup of the mechanics, the electronics and the software. We present the details of our motorized mirror realization, but want to stress that variations/adaptations of any kind are possible.

*Mechanical setup*

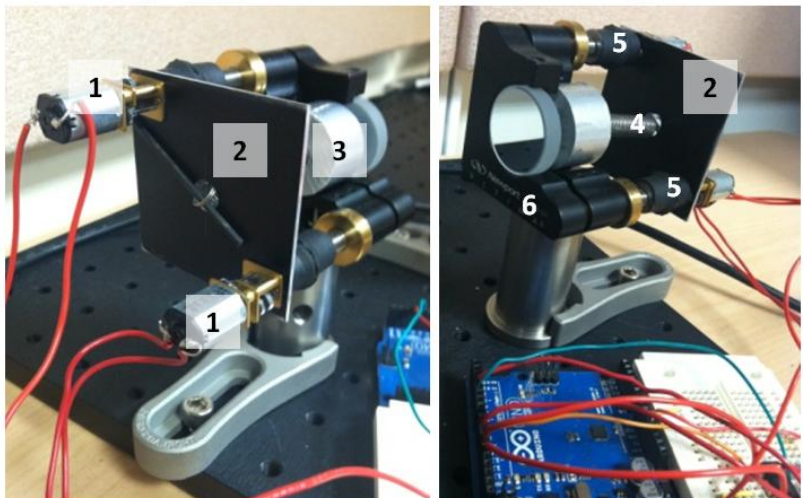

**Fig. 1**: Mechanical setup of the motorized mirror mount. The robotics motors (1) are attached to a sheet metal plate (2) (detail in Fig. 2a), whose details are given in Fig. 2a. The sheet metal plate is connected to the mirror mount via a spring (4) attached to an aluminum cylinder (3) (details in Fig. 2b) that is screwed to the mount. The motor axis is connected to the mirror fine thread screws via shaft couplers (5). One end of the shaft coupler screws directly on the motor shaft, the other end holds a hex key, which fits the fine thread actuator screws.

An overview of the mechanical setup is presented in Fig. 1. We chose to fit our setup on a commercial one-inch mirror Newport Ultima$^{TM}$ clear edge mirror mount with fine screws that fit a 5/64" diameter hex key (model #U100-A2H-NL). The design can easily adapted for a different mount by changing the motor mount positions on the sheet metal plate in Fig. 2a. The hobby robotics gearbox motors (1 in Fig.1) are metal, low current, brushed DC gearmotors from Pololu with extended motor shafts (product #1595)[10]) and provide enough torque for this setup due to their extreme gear ratio of 1000:1. Each of the two motors currently costs around 20 USD. The front plate of the motor has two tapped holes 9mm apart that fit M1.6 screws. The motors are attached to a sheet metal plate (2) shown in Fig. 2a. This plate is pushed onto the mirror mount by a spring (4,~1" long), attached by a securing pin through a central hole. The spring is fixed to the Al ring (3), which is in turn screwed onto the side of the mirror mount by an 8-32 screw. This construction allows for asymmetric elongation of the fine thread actuator screws of the mirror mount and avoids binding and damage to the motors/gearboxes. The motor shafts are screwed to 1/4'" shaft couplers, which have 5/64" hex rods on the other end. Since the hex rods do not quite fit many shaft couplers, flexible copper tape can be turned around the hex rods to provide a snug fit. Any play in the system between motor shaft and actuator screw will ultimately lead to backlash, which will be discussed later in this paper.

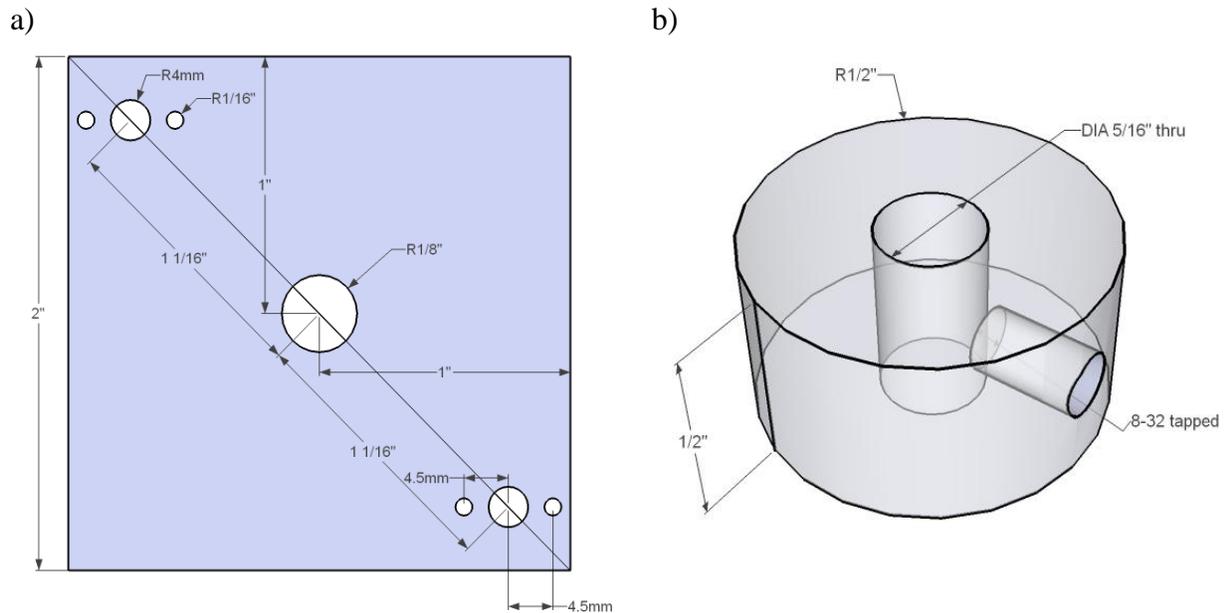

**Fig. 2**: a) Sheet metal plate holding the motors. The motor gearboxes have M1.6 threaded holes spaced a distance of 9mm, which allow installation on the sheet metal. The center hole in the three-hole pattern in the upper left and lower right is for the motor shaft. The spring fits through the center hole and is secured by a pin. The three-hole group in the upper left and lower right corner are identical. To make the graphic better readable, we called out different dimensions on the two groups.
b) Al ring that is fixed to the mirror mount via an 8-32 screw. The mirror mount frames have two holes fitting screws of this size. The center hole fits the spring that can be secured using a pin.

*Electronics setup*

Figure 3 illustrates the electronic layout of the driver circuit. The circuit was drawn using the free circuit diagramming software Fritzing [11]. In this figure, the black chip on the circuit board is the H-bridge (SN75441) from Texas Instruments [12]. The Arduino digital output pins connected to the H-bridge determine the speed and direction of the motors. The H-bridge uses three input pins for each motor. We demonstrate the driver operation for one motor. Two pins determine the direction of the motor. The motor turns one direction for pin 3 high and pin 6 low and vice versa for opposite signals. The speed is determined by the duty cycle of the pulse width modulation (PWM) signal connected to pin 1.

The Arduino microcontroller runs a program given in appendix A. The program can be uploaded on the Arduino using the software (IDE) and driver is available under arduino.cc. The Arduino IDE is available for many different computer operating systems. Our Arduino code is a modification of an existing H-bridge controller from the University of Massachusetts- Amherst [13]. The modified Arduino code in appendix A expects an eight character string input from a serial port. For this purpose, the Arduino board has a USB port acting as a serial connection. Each motor control is encoded by four characters, the first determining the direction (- being negative, 0 being positive). The remaining three characters correspond to the eight-bit range encoding the duty cycle of the Arduino PWM digital output (0 to 255). Eight element strings are thus transmitted and split into two strings (one for each motor), and each substring is used to control the H bridge, which in turn powers the two motors.

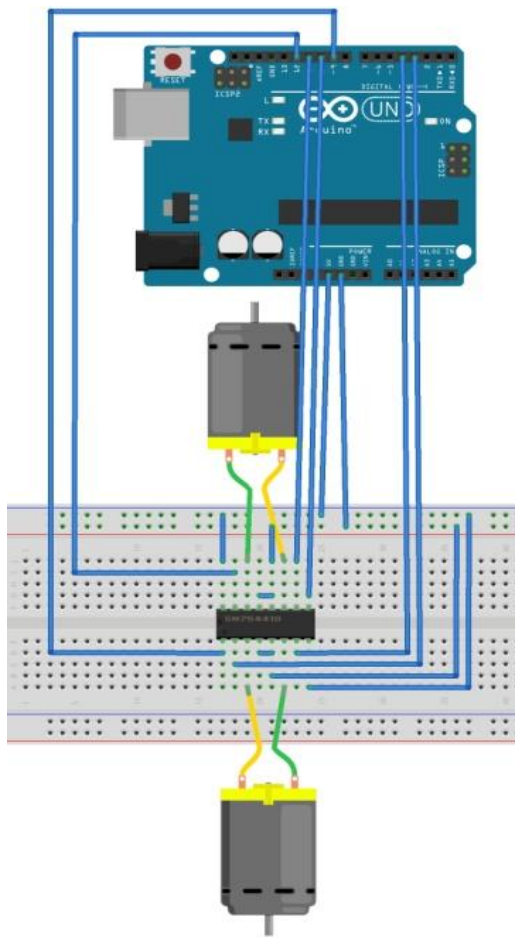

| Pin Output Number on H-Bridge | Connection | Pin Output Number on H-Bridge | Connection |
|---|---|---|---|
| 1 | Arduino Pin 9 | 9 | Arduino Pin 10 |
| 2 | Arduino Pin 3 | 10 | Arduino Pin 11 |
| 3 | Motor 1 Lead 1 | 11 | Motor 2 Lead 1 |
| 4 | Ground | 12 | Ground |
| 5 | Ground | 13 | Ground |
| 6 | Motor 2 Lead 2 | 14 | Motor 2 Lead 2 |
| 7 | Arduino Pin 4 | 15 | Arduino Pin 12 |
| 8 | +5 V | 16 | +5 V |

**Fig. 3:** Electronics setup. The gearbox motors (grey cylinders) are connected to an H bridge (SN754410 quadrupole half-h driver), which is controlled by an Arduino uno (alternatively

Arduino duemillanove) open hardware microcontroller board. The pinout of the H bridge and the Arduino is given in the table.

*Software*

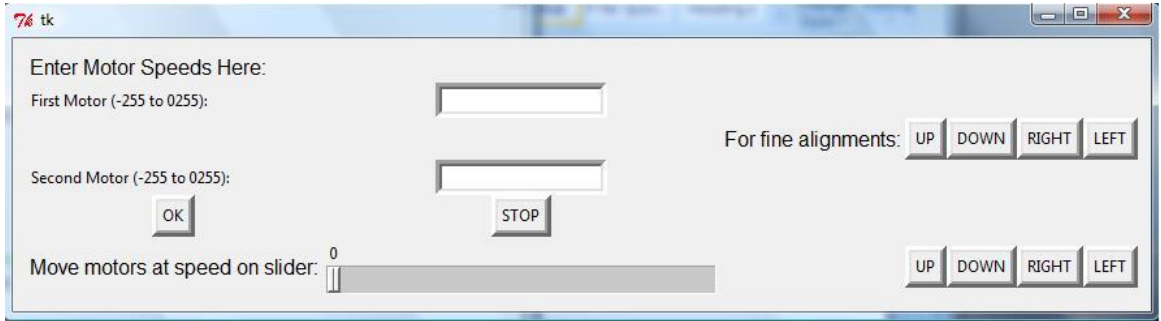

**Fig. 4**: Graphical user interface implemented by Python, as described in appendix B.

Any program that can use a serial interface can talk to the Arduino board. Among those are C, Matlab, and LabView. We use Python, which talks to the serial port via pyserial [14]. Python as well as its libraries are available as open source GPL compatible license. In addition, the TKinter library can be used for programming graphical user interfaces. The Python program translates the user input into an eight character string sent to the Arduino. Our code is shown in appendix B. The graphical user interface (GUI) is shown in Fig. 4. The three ways in which this GUI can be operated are:

1. Manually typing in the speeds for each motor
2. Dragging the slider out to a desired motor speed, then pressing and holding each of the buttons to the right of it to run one motor in the indicated direction at the speed on the slider
3. Pressing the buttons that can move the motors in small steps to finely orient the mirror mount. This is implemented by giving the motor full speed for a short time (called max_time in appendix B) only.

In addition, the stop button on the GUI stops the motors running completely during any of the aforementioned processes.

## III. Performance Test

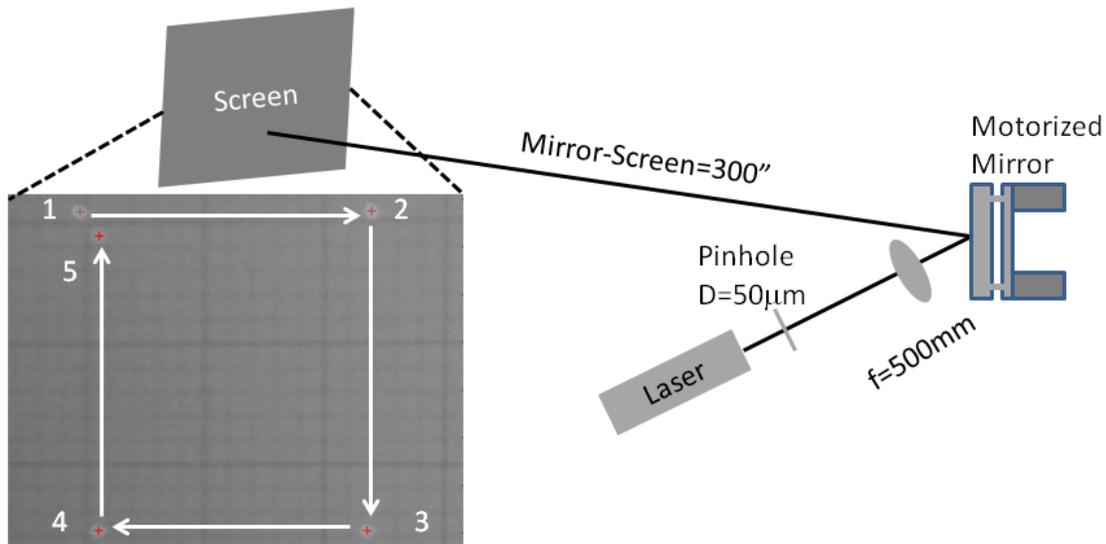

**Fig. 5**: Setup and result of the motorized mirror test. A laser illuminated pinhole is imaged on a far screen by a lens via the motor controlled folding mirror. The mirror was scanned 20 steps in a square. One step corresponds to 0.43 and 0.47 mrad in the two orthogonal directions. The fact that the square does not close is due to backlash of the gearbox, shaft coupler and hexrod assembly.

With the circuitry completed and the motorized mirror mount built, the system needed to be assessed to ensure functionality. Each feature on the GUI was tested at various settings (i.e. speed, length the buttons were held down for, different inputs into the manual input boxes), and all worked as anticipated. In order to quantify the current setup, we chose to measure the angular deviation of a laser beam impinging on the computer controlled mirror. The setup is shown in Fig. 5a. A pinhole, which is illuminated by a laser pointer, is imaged on a screen via the motor controlled mirror. The screen is 300" away from the mirror, thus small angles can be measured with high accuracy. The screen is photographed via a webcam and several images are overlaid, as shown in the inset. The motorized mirror is moved to the position 1 by actuating the beam right and down, thus those two directions have no backlash. We move the beam from position 1 to position 2 with 20 steps to the right (in the GUI). From 2 to 3 we go 20 steps down. Changing the direction to the left or upwards results in a backlash. The following twenty steps left (to 4) and 20 steps up (to 5) thus result in a spot that is not overlapped with position 1. Finding the centroid of the laser spot (red crosses) shows that 1 step, as currently implemented, results in 0.43 mrad x motion and 0.47 mrad y motion. Even smaller steps are possible by either reducing the speed of the step or alternatively the time that the motor actuates (currently 0.2 sec). The backlash in the x and y directions is determined by the distance between spots 1 and 5 which is 0.64 mrad.in the x direction and 0.7 mrad in the y direction. The backlash results from some play in the gearbox and coupler-hexrod assembly. Once it is overcome (takes about 1 step) the motion is fully following the controls. Several tests confirmed that the backlash is repeatable and thus an automatic backlash correction is possible.

## IV. Conclusion

With all features working as anticipated, the motorized mirror mount can now be used in undergraduate research laboratories and can also be applied in general optics laboratory settings. Possible extensions of the mount can include stop switches that stop the motion once the fine thread screw has reached a certain maximal or minimal position. In addition optical or magnetic encoders can be used to absolutely calibrate steps, if needed. One extension of this project includes a feedback on webcams. By fixing the laser beam position on two target points using two webcams and two motorized mirror mounts, fully automatic beam-pointing stabilization can be implemented. Due to use of consumer parts, this costs a fraction of comparable commercial systems. In the design of the current motor controlled setup, we have not at all optimized for high speed. This would require a totally different approach using piezo crystals as actuators.

## V. Acknowledgments

M. Gopalakrishnan would also like to thank the science undergraduate laboratory internship (SULI) program at SLAC National Accelerator Laboratory sponsored by the Office of Science, U. S. Department of Energy. This work was supported by the AMOS program within the Chemical Sciences, Geosciences, and Biosciences Division of the Office of Basic Energy Sciences, Office of Science, U.S. Department of Energy. M. Gühr acknowledges funding via the Office of Science Early Career Research Program through the Office of Basic Energy Sciences, U.S. Department of Energy. We acknowledge fruitful discussions with B. K. McFarland and Matthew Ware.

## VI. Appendix A: Arduino Code (for future supplementary online information)

```
String readString,motor1,motor2;

int motor1Pin = 3; // H-bridge pin2
int motor2Pin = 4; // H-bridge pin 7
int motor3Pin = 11; // H-bridge pin 10
int motor4Pin = 12; // H-bridge pin 15
int speedPin1 = 9; // H-bridge enable pin 1
int speedPin2 = 10; // H-bridge enable pin 9
int speed = 0; //Set start speed to be 0

void setup() {
// Load serial monitor
Serial.begin (9600);

// set all the other pins you're using as outputs:
pinMode(motor1Pin, OUTPUT);
pinMode(motor2Pin, OUTPUT);
pinMode(motor3Pin, OUTPUT);
pinMode(motor4Pin, OUTPUT);
pinMode(speedPin1, OUTPUT);
pinMode(speedPin2, OUTPUT);
}

void loop(){
  readString="";
  while (Serial.available()){
    delay(5);
    if (Serial.available() > 0) {
      char c = Serial.read();
      readString+= c;
    }
  }
  Serial.flush();
  if (readString.length() >0) {
   Serial.println(readString); //Read string inputted to serial monitor

   //Split string into two and send one part of the string to each motor
   motor1=readString.substring(0,4);
   motor2=readString.substring(4,8);

  //Convert strings to integers so the motors can read the numbers
   int n1=motor1.toInt();
   int n2=motor2.toInt();

//Print the integers written to each motor
   Serial.println(n1);
   Serial.println(n2);

// if the switch is high, motor 1 will turn on one direction:
if (n1>0) {
digitalWrite(motor1Pin, LOW); // set leg 1 of the H-bridge low
digitalWrite(motor2Pin, HIGH); // set leg 2 of the H-bridge high
speed = abs(n1);
analogWrite (speedPin1, speed);
Serial.println(speed);

delay (50);

}

// if the switch is low, motor 1 will turn in the other direction:
else {
digitalWrite(motor1Pin, HIGH); // set leg 1 of the H-bridge high
digitalWrite(motor2Pin, LOW); // set leg 2 of the H-bridge low
speed = abs(n1);
analogWrite (speedPin1, speed);
Serial.println(speed);
delay (50);
}

// if the switch is low, motor 2 will turn in the other direction:
if (n2>0) {
digitalWrite(motor3Pin, LOW); // set leg 1 of the H-bridge low
digitalWrite(motor4Pin, HIGH); // set leg 2 of the H-bridge high
speed = abs(n2);
analogWrite (speedPin2, speed);
Serial.println(speed);
delay (50);
}
// if the switch is low, motor 2 will turn in the other direction:
else {
digitalWrite(motor3Pin, HIGH); // set leg 1 of the H-bridge high
digitalWrite(motor4Pin, LOW); // set leg 2 of the H-bridge low
speed = abs(n2);
analogWrite (speedPin2, speed);
Serial.println(speed);
delay (50);
}
}
delay(2);
}
```

## VII. Appendix B: Python Code (the code extends over the complete left column over the next couple of pages, then follows the code on the right colunm)

```
### SETTING UP ###
from __future__ import print_function
from Tkinter import *
from tkFont import Font
import RPi.GPIO as GPIO
from subprocess import call
import time
from time import sleep

###SLIDER CODE###
        global var
        var = IntVar()

        slider=Scale(top,orient=HORIZONTAL,length=300,width=20,slide
rlength=10,from_=0,to=255, variable = var).grid(row=9,column=1)

        def stopi(self):
```

```python
import serial
global ser
ser=serial.Serial(COM4, timeout=1)

###DEFINING STEPI FUNCTION###

def stepi(str):
    time.sleep(0.1)
    print(ser.write(str))
    print(str)
    time.sleep(0.1)

###STARTING A CLASS###

class MyDialog:
    def __init__(self, parent):
        self.parent = parent
        self.mouse_pressed = False

###OPENING SERIAL MONITOR###

        global ser
        ser=serial.Serial(COM4,9600, timeout=1)
        time.sleep(3)
        top = self.top = Toplevel(parent,bd=10)

###LABELS###

        Label(top,text="Enter Motor Speeds Here:",font='bold').grid(row=0,sticky=W)
        Label(top,text="First Motor (-255 to 0255):",bd=3).grid(row=2,sticky=W)
        Label(top,text="Second Motor (-255 to 0255):",bd=3).grid(row=4,sticky=W)
        Label(top,text="For fine alignments:",font='bold').grid(row=3,column=7)
        Label(top,text="Move motors at speed on slider:",font='bold').grid(row=9,column=0)

###SETTING ENTRIES, "OK" AND "STOP" BUTTONS###

        self.e = Entry(top,bd=5)
        self.e.grid(row=2,column=1)

        self.f = Entry(top,bd=5)
        self.f.grid(row=4,column=1)

        b = Button(top, text="OK",bd=5,command=self.ok)
        b.grid(row=7,column=0)

        c = Button(top, text="STOP",bd=5,command=self.stop)
        c.grid(row=7,column=1)
widg4.grid(row=9,column=12)

    def dec4(self):
        m=var.get()
        m1=str(m)
        print("-%03d"%m)
        n='0000'
        stepi(n + "-%03d"%m);

        stepi('00000000')
        print('stopped')

        widg1 = Button(top,text="UP",bd=5)
        widg1.grid(row=9,column=9)

    def inc3(self):
        m=var.get()
        m1=str(m)
        print("'%04d'"%m)
        n='0000'
        stepi("%04d"%m + n);

        widg1.bind('<Button-1>',inc3)
        time.sleep(0.1)
        widg1.bind('<ButtonRelease-1>',stopi)

        widg2 = Button(top,text="DOWN",bd=5)
        widg2.grid(row=9,column=10)

    def dec3(self):
        m=var.get()
        m1=str(m)
        print("-%03d"%m)
        n='0000'
        stepi("-%03d"%m + n);

        widg2.bind('<Button-1>',dec3)
        time.sleep(0.1)
        widg2.bind('<ButtonRelease-1>',stopi)

        widg3 = Button(top,text="RIGHT",bd=5)
        widg3.grid(row=9,column=11)

    def inc4(self):
        m=var.get()
        m1=str(m)
        print("'%04d'"%m)
        n='0000'
        stepi(n + "%04d"%m);

        widg3.bind('<Button-1>',inc4)
        time.sleep(0.1)
        widg3.bind('<ButtonRelease-1>',stopi)

        widg4 = Button(top,text="LEFT",bd=5)

###DEFINING BUTTONS FOR FINE TUNING MIRROR MOUNT ORIENTATION###

    def inc(self):
        max_time = 0.2
        print(time.time())
        start_time = time.time()
        print(start_time)
        stepi('-2550000');
```

```python
    widg4.bind('<Button-1>',dec4)
    time.sleep(0.1)
    widg4.bind('<ButtonRelease-1>',stopi)

###BUTTONS FOR FINE TUNING ORIENTATION OF MIRROR MOUNT###

    inc = Button(top,text="UP",bd=5,command=self.inc)
    inc.grid(row=3,column=9)

    dec = Button(top,text="DOWN",bd=5,command=self.dec)
    dec.grid(row=3,column=10)

    inc2 = Button(top,text="RIGHT",bd=5,command=self.inc2)
    inc2.grid(row=3,column=11)

    dec2 = Button(top,text="LEFT",bd=5,command=self.dec2)
    dec2.grid(row=3,column=12)

###DEFINING "OK" FUNCTION###
    def ok(self):
        m=self.e.get()
        print(m)
        n=self.f.get()
        print(n)
        stepi(m + n);
        print(m)
        print(n)

###DEFINING "STOP" FUNCTION###

    def stop(self):
        stepi('00000000');

        while (time.time() - start_time) < max_time:
            print(time.time() - start_time)
        stepi('00000000');

    def dec(self):
        max_time = 0.2
        print(time.time())
        start_time = time.time()
        print(start_time)
        stepi('02550000');
        while (time.time() - start_time) < max_time:
            print(time.time() - start_time)
        stepi('00000000');

    def inc2(self):
        max_time = 0.2
        print(time.time())
        start_time = time.time()
        print(start_time)
        stepi('00000255');
        while (time.time() - start_time) < max_time:
            print(time.time() - start_time)
        stepi('00000000');

    def dec2(self):
        max_time = 0.2
        print(time.time())
        start_time = time.time()
        print(start_time)
        stepi('0000-255');
        while (time.time() - start_time) < max_time:
            print(time.time() - start_time)
        stepi('00000000');

###END—CLOSE SERIAL MONITOR###

root = Tk()
try:
    d = MyDialog(root)
    root.wait_window(d.top)
except IOError:
    ser.close()
```